\documentclass[preprint]{ptephy_v1}

\preprintnumber{YITP-24-168, KUNS-3030} 
\usepackage{hyperref}

\usepackage{bm}
\usepackage[all]{xy}
\usepackage{tcolorbox}
\usepackage{tcolorbox}
\tcbuselibrary{breakable, skins, theorems}
\usepackage{mathrsfs}

\newtheorem{theorem}{Theorem}

\newtheorem{definition}{Definition}

\newtheorem*{theorem*}{Theorem}

\numberwithin{equation}{subsection}
\numberwithin{definition}{subsection}
\numberwithin{theorem}{subsection}
\numberwithin{remark}{subsection}
\numberwithin{conjecture}{subsection}

\usepackage[utf8]{inputenc}

\DeclareFontFamily{U}{min}{}

\DeclareFontShape{U}{min}{m}{n}{<-> udmj30}{}

\newcommand\yo{\!\text{\usefont{U}{min}{m}{n}\symbol{'207}}\!}

\begin{document}

\title{States and IR divergences in factorization algebras}


\author{Masashi Kawahira}
\affil{Yukawa Institute for Theoretical Physics, Kyoto University, Kyoto 606-8502 Japan \email{masashi.kawahira@yukawa.kyoto-u.ac.jp}}

\author[2]{Tomohiro Shigemura}
\affil{Department of Physics, Kyoto University, Kyoto 606-8502, Japan \email{shigemura@gauge.scphys.kyoto-u.ac.jp}}

\begin{abstract}
In field theory, one can consider a variety of states.
Within the framework of factorization algebras, one typically works with the natural augmentation state $\langle-\rangle_{\rm aug}$. 
In physics, however, other states arise naturally, such as the compactification state $\langle-\rangle_{\rm cptf}$ or the Schwartz state $\langle-\rangle_{\rm Sch}$, defined by imposing Schwartz boundary conditions.
At first sight, the relation among these three states is not obvious.

This paper gives a definition of the compactification state in factorization algebras and provides a method for handling infrared divergences in the massless theory.
We then prove that the three states are equivalent in both the massive and massless cases.
\end{abstract}

\subjectindex{A12, A13, B05, B33 B34, B39}
\maketitle
\tableofcontents
\section{Introduction and summary}

Quantum field theory (QFT) is a central framework in modern theoretical physics.
It successfully describes a wide range of physical phenomena, and it has also inspired numerous deep conjectures in mathematics.
Despite its remarkable success, however, a complete conceptual and mathematical understanding of QFT is still lacking.

Over the past decade, Kevin Costello and Owen Gwilliam have developed a new formulation of quantum field theory based on factorization algebras\cite{Costello:2016vjw,Costello:2021jvx}.
This framework is particularly well suited to perturbative field theories.
Its main advantage is that it clarifies many fundamental concepts in QFT, such as path integrals and operator products.

In this paper, we aim to clarify the origin of infrared divergences (IR divergences) in massless free scalar theories.
IR divergences have been studied for many decades\cite{Bloch:1937pw}, and the topic has recently attracted renewed attention\cite{Strominger:2017zoo}.
We show that, within the framework of factorization algebras, IR effects can be treated without encountering explicit divergences.

In order to discuss IR divergences in a mathematically rigorous way, we organize the construction of states in factorization algebras.
We focus on three states, which we refer to as {\it the natural augmentation state, the compactification state}, and the {\it Schwartz state}. 

We consider observables supported in an open subset $U\subset \mathbb{R}^d$ , which we denote by $\mathcal{O}_U\in{\rm Obs}(U):={\rm Sym}(C_{\rm c}^\infty(U))$. 
In the framework of factorization algebras, taking cohomology and choosing a state yields expectation values\footnote{Note that $\mathcal{O}_U$ is an observable supported in the subset $U$, and $[-]_U$ denotes taking cohomology on $U$. Precise definitions of cohomology and states are given in Section \ref{sec:Brief_review}.}.
\begin{align}
    H^0{\rm Obs}^{\rm cl}(U)\ni [\mathcal{O}_U]_U\mapsto \langle [\mathcal{O}_U]_U\rangle \in\mathbb{C}
\end{align}

We now briefly summarize the three states of interest:
\begin{itemize}
    \item Natural augmentation state 
    $\langle-\rangle_{\rm aug}$
    \item Schwartz state
    $\langle-\rangle_{\rm Sch}$
    \item Compactification state
    $\langle-\rangle_{\rm cptf}$
\end{itemize}
\noindent{}
The states $\langle-\rangle_{\rm aug}$ and $\langle-\rangle_{\rm Sch}$ were introduced by Costello and Gwilliam\cite{Costello:2016vjw}, who proved that they are equivalent.
The natural augmentation state $\langle-\rangle_{\rm aug}$ is well defined in both the massive and massless cases, whereas the Schwartz state $\langle-\rangle_{\rm Sch}$ is defined only in the massive case. 
In the massless case, the construction fails because of an infrared (IR) divergence.
One of the main purposes of this paper is to show how to remove this IR divergence and thereby define the Schwartz state also in the massless case.

In addition, we define the compactification state $\langle-\rangle_{\rm cptf}$, which captures the notion of compactification in the factorization algebra framework.
This state is well defined in the massive case.
In the massless case, however, one must first eliminate the IR divergence in order to define $\langle-\rangle_{\rm cptf}$.
Furthermore, we prove that all three states are equivalent in both the massive and massless cases.

The remainder of this paper is structured as follows.
Section 2 reviews the formulation developed by Costello and Gwilliam.
The natural augmentation state is examined in Section 3.
Section 4 presents the compactification state and provides a detailed treatment of the associated IR divergences.
Subsequently, Section 5 first reviews the Schwartz state for the massive case, followed by its extension to the massless case, including a discussion on handling IR divergences.
The equivalence of the aforementioned states is established in Section 6.
Finally, Section 7 provides discussion.

\section{Brief review of factorization algebra of free real scalar theory}
\label{sec:Brief_review}

\subsection{Observable algebra}

In this paper, we discuss path integral formulation in a mathematical manner.
In contrast to operator formulation, observables are not operators but functionals with compact support.
We define observable algebra as follows.

\begin{tcolorbox}[colframe=red,colback=red!3!]
\begin{definition}{\rm (K. Costello, O. Gwilliam\cite[\S 2.2]{Costello:2016vjw})}\\
The observable algebra ${\rm Obs}(U)$\footnotemark on an open subset $U\subset M$ is defined as\begin{align}
    {\rm Obs}(U):={\rm Sym}(C_{\rm c}^\infty(U,\mathbb{C}))
\end{align} and the elements are called observables on $U$.
\end{definition}
 \end{tcolorbox}
\footnotetext{Precisely, a completion is required to make $\mathrm{Obs}(U)$ a topological vector space. For details, see\cite{Costello:2016vjw}.}

The motivation for the above definition is that observables can be regarded as a polynomial of functions with compact support:
$\mathcal{O}=c+f+f_1*f_2+\cdots$,
where $*$ is a formal symmetric product, $c\in\mathbb{C}$, and $f,f_1,f_2,\cdots\in C_{\rm c}^\infty(U,\mathbb{C})$. 
The observable $\mathcal{O}$ acts on the field $\Phi$ as
\begin{align}
    \mathcal{O}:\Phi\mapsto
    c
    +\int_M f\Phi
    +\int_M f_1\Phi \int_M f_2\Phi
    +\cdots.
\end{align}

\subsection{Classical derived observable space}

To formulate path integrals rigorously, we introduce the concept of derived observable algebras. The term ``derived'' signifies the incorporation of a degree structure into the observable space ${\rm Obs}(U)$. This formulation originates from the work of Batalin and Vilkovisky\cite{Batalin:1981jr, Batalin:1983ggl}.

\begin{align}
    &C_{\rm c}^\infty(U)^{0}:=(C_{\rm c}^\infty(U,\mathbb{C}),0),\\
    &C_{\rm c}^\infty(U)^{-1}:=(C_{\rm c}^\infty(U,\mathbb{C}),-1).
\end{align}
The first is the \textit{degree-0 linear observable algebra}, and the second is the \textit{degree-$(-1)$ linear observable algebra}, which is often identified with the \textit{anti-field space} in physics.
For simplicity, we denote $(f,0)\in C_{\rm c}^\infty(U)^{0}$ and $(f,-1)\in C_{\rm c}^\infty(U)^{-1}$ as $f$ and $f^\star$.
The symmetric product $*$ is defined for them as
\begin{align}
    a*b=(-1)^{|a||b|}b*a\label{eq:*}.
\end{align}

Furthermore, we define a degree-$(+1)$ operator $\Delta^{\rm cl}$ as follows.

\begin{tcolorbox}[colframe=red,colback=red!3!]
\begin{definition}{\rm (K. Costello, O. Gwilliam\cite[\S 4.2]{Costello:2016vjw})}\label{def:cl_BV_op}\\
     The classical Batalin-Vilkovisky operator $\Delta^{\rm cl}$ is a map 
     \begin{align}
      C_{\rm c}^\infty(U)^{-1}\ni f^\star\mapsto -(-\Delta+m^2)f\in C_{\rm c}^\infty(U)^{0}.   
     \end{align}
     Hence $\Delta^{\rm cl}$ has a degree-$(+1)$.
\end{definition}    
\end{tcolorbox}
\noindent{}The explicit form of $\Delta^{\rm cl}$ is theory-dependent. As we restrict our attention to free theories, the definition provided above is employed. 
To incorporate interactions, the operator $\Delta^{\rm cl}$ must be deformed by adding additional terms.
The classical derived observable algebra is subsequently defined using the classical Batalin-Vilkovisky operator.

\begin{tcolorbox}[colframe=red,colback=red!3!]
\begin{definition}{\rm (K. Costello, O. Gwilliam\cite[\S 4.2]{Costello:2016vjw})}\\
    The classical derived observable algebra ${\rm Obs}^{\rm cl}(U)$ is defined as
    \begin{align}
        {\rm Obs}^{\rm cl}(U)
        :=
        {\rm Sym}\left(C_{\rm c}^\infty(U)^{-1}\xrightarrow{\Delta^{\rm cl}} C_{\rm c}^\infty(U)^{0}\right)
    \end{align}
    where $\Delta^{\rm cl}$ is a classical Batalin-Vilkovisky operator which is defined in Definition \ref{def:cl_BV_op}.
\end{definition}    
\end{tcolorbox}

By (\ref{eq:*}), we can rewrite the classical derived observable space ${\rm Obs}^{\rm cl}(U)$.
\begin{align}
   &{\rm Obs}^{\rm cl}(U)= 
   \notag\\
   &\bigg(\cdots 
   \xrightarrow{\Delta^{\rm cl}}\bigwedge^2 C_{\rm c}^\infty(U)^{-1}* {\rm Sym}\left(C_{\rm c}^\infty(U)^{0}\right)
   \xrightarrow{\Delta^{\rm cl}}C_{\rm c}^\infty(U)^{-1}* {\rm Sym}\left(C_{\rm c}^\infty(U)^{0}\right)
   \xrightarrow{\Delta^{\rm cl}}{\rm Sym}\left(C_{\rm c}^\infty(U)^{0}\right)\bigg)
\end{align}
This complex is known as the \textit{classical Batalin–Vilkovisky complex}, and its cohomology is referred to as the \textit{classical Batalin–Vilkovisky cohomology}
 \begin{align}
   H^*\left({\rm Obs}^{\rm cl}(U)\right).
 \end{align}

Let us consider the physical meanings of the cohomology $H^0\left({\rm Obs}^{\rm cl}(U)\right)$.
Take degree-0 observables $\mathcal{O}_1,\mathcal{O}_2$ and assume that these are equivalent as the cohomology class, i.e., $\exists X\ {\rm s.t.}$
\begin{align}
    \mathcal{O}_2-\mathcal{O}_1
    =
    \Delta^{\rm cl}X.
\end{align}
Let $\Phi_{\rm cl}$ be a solution of the equation of motion $(-\Delta+m^2)\Phi=0$. We have
\begin{align}
    \mathcal{O}_2(\Phi_{\rm cl})-\mathcal{O}_1(\Phi_{\rm cl})
    &=
    \Delta^{\rm cl}X(\Phi_{\rm cl})\notag\\
    &=
    0.
\end{align}
Consequently, $H^0\left({\rm Obs}^{\rm cl}(U)\right)$ can be interpreted as the on-shell evaluation of observables. However, the resulting bracket $[\mathcal{O}_1](=[\mathcal{O}_2])$ is not a scalar value. To address this, we define a map, denoted as the \textit{state}, $\langle-\rangle$.
\begin{tcolorbox}[colframe=red,colback=red!3!]
\begin{definition}{\rm (K. Costello, O. Gwilliam\cite[\S 4.9]{Costello:2016vjw})}\\
    A state $\langle-\rangle$ is a smooth\footnotemark map:
    \begin{align}
        \langle-\rangle:
        H^0\left({\rm Obs}^{\rm cl}(M)\right)
        \to 
        \mathbb{C}.
    \end{align}
\end{definition}
\end{tcolorbox}
\footnotetext{Here, the definition of \textit{smooth} is based on the notion of a smooth set. A brief review of smooth sets is given in Appendix \ref{sec:smooth_set}.}

\section{Natural augmentation state}

\subsection{Massive and massless cases}

By definition, the classical derived observable space ${\rm Obs}^{\rm cl}(\mathbb{R}^d)$ has a natural augmentation map. Costello and Gwilliam use this to define a state.
\begin{tcolorbox}[colframe=red,colback=red!3!]
\begin{definition}
{\rm (K. Costello, O. Gwilliam\cite[\S 4.9]{Costello:2016vjw})}\\
    We have the natural augmentation map:
    \begin{align}
        \begin{array}{ccc}
         {\rm Obs}^{\rm cl}(\mathbb{R}^d)^{0}
        &\longrightarrow
        &\mathbb{C}\\
         \rotatebox{90}{$\in$}
        &
        &\rotatebox{90}{$\in$}\\
        c+f+f_1*f_2+\cdots
        &\longmapsto
        &c
        \end{array}
    \end{align}
    where $c\in\mathbb{R}$ and $f,f_1,f_2,\cdots\in C_{\rm c}^\infty(U)$.
    This induces
    \begin{align}
        \langle-\rangle_{\rm aug}:
        H^0({\rm Obs}^{\rm cl}(\mathbb{R}^d))
        \to
        \mathbb{C}.
    \end{align}
    We call it the natural augmentation state.
\end{definition}
\end{tcolorbox}
Let us explain why 
the natural augmentation state is well defined.
Here, the cohomology is
\begin{align}
    H^0({\rm Obs}^{\rm cl}(\mathbb{R}^d))
    =
    \frac
    {{\rm Obs}^{\rm cl}(\mathbb{R}^d))^{0}}
    {\Delta^{\rm cl}
    ({\rm Obs}^{\rm cl}(\mathbb{R}^d)^{-1})}.
\end{align}
To establish the well-definedness of $\langle-\rangle_{\rm aug}$, we demonstrate that for any $X\in {\rm Obs}^{\rm cl}(\mathbb{R}^d)^{-1}$,
\begin{align}
    \Delta^{\rm cl}X
    \notin
    (\mathbb{C}\setminus\{0\})
    \subset
    {\rm Obs}^{\rm cl}(\mathbb{R}^d)^{0}.
\end{align}
Decompose $X$ as $X=P*f^\star$ with $P$ in degree-$0$ and $f^\star$ in degree-$(-1)$.
Applying $\Delta^{\rm cl}$ gives
\begin{align}
    \Delta^{\rm cl}X
    =
    P*(\Delta^{\rm cl}f^\star),
\end{align}
since  $\Delta^{\rm cl}$ acts trivially on $P$.
For $\Delta^{\rm cl}X$ to be an element of $\mathbb{C}$, both $P$ and $\Delta^{\rm cl}f^\star$ must be constant functions.
However, by definition, $\Delta^{\rm cl}f^\star$ belongs to the space of compactly supported smooth functions in degree-$0$:
\begin{align}
    \Delta^{\rm cl}f^\star
    \in
    C_{\rm c}^\infty(\mathbb{R}^d)^0.
\end{align}
Since the only constant function contained in $C_{\rm c}^\infty(\mathbb{R}^d)^0$ is the zero function, we must have $\Delta^{\rm cl}f^\star = 0$. Consequently, this ensures that the term $\langle-\rangle_{\rm aug}$ is well-defined.

\section{Compactification state}
\subsection{Massive case}

In order to consider compactification, locality of the observables plays an essential role.
Therefore, we assume that the open subset $U$ is bounded, which we refer to {\it the locality condition}.
Under this condition, we have an inclusion map
\begin{align}
    i:{\rm Obs}^{\rm cl}(U)\to 
    {\rm Obs}^{\rm cl}(T^d)
\end{align}
for sufficiently large torus $T^d$.
And hence
\begin{align}
    i:H^*{\rm Obs}^{\rm cl}(U)\to 
    H^*{\rm Obs}^{\rm cl}(T^d)
\end{align}
holds. 
Strictly speaking, these two maps are distinct; however, we denote both by $i$ for simplicity.
\begin{tcolorbox}[colframe=blue,colback=blue!3!]
\begin{theorem}
{\rm (K. Costello, O. Gwilliam\cite[\S 2.5]{Costello:2016vjw})}\\
    In the massive case,
    \begin{align}
    H^n{\rm Obs}^{\rm cl}(T^d)
    =
    \left\{
    \begin{array}{ll}
    \mathbb{C} & (n=0) \\
    0 & ({\rm otherwise})
    \end{array}
    \right.
\end{align}
\end{theorem}
\end{tcolorbox}
\begin{proof}
    \begin{align}
        A:=
        \left(C_{\rm c}^\infty\left(T^d\right)^{-1}\xrightarrow{\Delta^{\rm cl}}C_{\rm c}^\infty\left(T^d\right)^0\right)
    \end{align}
    This is an isomorphism. Thus $H^*(A)=0$, then $H^0({\rm Sym}(A))=\mathbb{C}$ and $H^{n\le -1}({\rm Sym}(A))=0$
\end{proof}

Then the inclusion map $i$ gives a state $\langle-\rangle_{\rm cptf}$.
\footnote{The abbreviation ``cptf" stands for ``compactification."}
\begin{tcolorbox}[colframe=red,colback=red!3!]
\begin{definition}\ \\
    The compactification state $\langle-\rangle_{\rm cptf}$ is the smooth map:
    \begin{align}
    \langle-\rangle_{\rm cptf}:
        H^0{\rm Obs}^{\rm cl}(U)
    \to
    H^0{\rm Obs}^{\rm cl}(T^d)=\mathbb{C}.
    \end{align}
\end{definition}
\end{tcolorbox}

\subsection{Massless case}
\begin{tcolorbox}[colframe=blue,colback=blue!3!]
\begin{theorem}\ \\
    In the massless case,
    \begin{align}
    H^*{\rm Obs}^{\rm cl}(T^d)
    =
    \mathbb{C}[q,r]/(r^2)
    \end{align}
    where $q$ is a generator with degree-$0$ and $r$ is a generator with degree-$(-1)$.
\end{theorem}
\end{tcolorbox}
\begin{proof}
    We show a quasi-isomorphism:
    \begin{align}
        A\xrightarrow{\pi} B
    \end{align}
    where
    \begin{align}
        A=\left(C_{\rm c}^\infty\left(T^d\right)^{-1}\xrightarrow{\Delta^{\rm cl}}C_{\rm c}^\infty\left(T^d\right)^0\right),\ 
        B=(\mathbb{C}\xrightarrow{0}\mathbb{C}).
    \end{align}
    We denote the bases of the two $\mathbb{C}$'s in the complex $B$ as $r,q$, then we have $B=(\mathbb{C}r \xrightarrow{0} \mathbb{C}q)$.
    Once the quasi-isomorphism between $A$ and $B$ is proven, ${\rm Sym}(A)$ and ${\rm Sym}(B)$ are also quasi-isomorphic\footnote{This is because taking the symmetric product commutes with taking cohomology.}. Thus,
    \begin{align}
        H^*{\rm Obs}^{\rm cl}(T^d) = H^* {\rm Sym} (A) \simeq H^* {\rm Sym} (B) \simeq \mathbb{C}[q,r]/(r^2).
    \end{align}

    First of all, we show the following commutative diagram:
    \begin{align}
        \xymatrix{
    C_{\rm c}^\infty\left(T^d\right)^{-1} \ar[r]^{\Delta^{\rm cl}} \ar[d]_{\pi^{-1}} & C_{\rm c}^\infty\left(T^d\right)^0 \ar[d]_{\pi^0} \\
    \mathbb{C} \ar[r]^{0} & \mathbb{C}
    }
    \end{align}
    where $\pi^{-1}$ is defined as  
   \begin{align}
       \pi^{-1}(f^{\star})
       :=
       r\int_{T^d}{\rm d}x_1 \cdots {\rm d}x_d\ f(x_1,\cdots,x_d)
       =
       r\int_{0}^{2\pi} {\rm d}x_1
       \cdots
       \int_{0}^{2\pi} {\rm d}x_d
       \ f(x_1,\cdots,x_d)
   \end{align}
   for $f^{\star}\in C_{\rm c}^\infty\left(T^d\right)^{-1}$, 
   and $\pi^{0}$ is defined as
   \begin{align}
       \pi^{0}(g)
       :=
       q\int_{T^d}{\rm d}x_1\cdots{\rm d}x_d\ g(x_1,\cdots,x_d)
       =
       q\int_{0}^{2\pi} {\rm d}x_1
       \cdots
       \int_{0}^{2\pi} {\rm d}x_d
       \ g(x_1,\cdots,x_d)
   \end{align}
   for $g\in C_{\rm c}^\infty\left(T^d\right)^0$.
    We can easily check that
    $\pi^0(g)=0$ holds if $g=\Delta^{\rm cl}f^{\star}.$

Second, we show $H^0(A)=H^0(B)$. By definition,
\begin{align}
    H^0(A)
    =
    \frac{C_{\rm c}^\infty(T^d)^0}
    {{\rm im}(\Delta^{\rm cl})}.
\end{align}
We claim that ${\rm im}(\Delta^{\rm cl})={\rm ker}(\pi^0)$. Once this is established, it follows immediately that $H^0(A)=H^0(B)$. Obviously,
\begin{align}
    {\rm im}(\Delta^{\rm cl})\subset {\rm ker}(\pi^0).
\end{align}
Next, we verify that ${\rm im}(\Delta^{\rm cl})\supset {\rm ker}(\pi^0)$.
Take $f\in {\rm ker}(\pi^0)$. $f$ satisfies
\begin{align}
    \int_{T^d} {\rm d}x_1 \cdots {\rm d}x_d\ f(x_1,\cdots,x_d)=0.\label{eq:int=0}
\end{align}
Then we define functions $\tilde{f}$ and $\tilde{\tilde{f}}$ as
\begin{align}
    &\tilde{f}(x_1,\cdots,x_d)
    := \sum_{i=1}^{d} \int_{0}^{2\pi}{\rm d}y_1 \cdots \int_{0}^{x_i}{\rm d}y_i \cdots \int_{0}^{2\pi}{\rm d}y_d\ f(y_1,\cdots,y_d),\\
    &\tilde{\tilde{f}}(x_1,\cdots,x_d)
    :=
    \tilde{f}(x_1,\cdots,x_d)-\frac{1}{(2\pi)^d}\int_{T^d}
    {\rm d}y_1\cdots{\rm d}y_d\ \tilde{f}(y_1,\cdots,y_d).
\end{align}
$\tilde{f}$ and $\tilde{\tilde{f}}$ are in $C_{\rm c}^\infty(T^d)$ because of (\ref{eq:int=0}). And we have
\begin{align}
    \int_{T^d}{\rm d}x\ \tilde{\tilde{f}}(x_1,\cdots,x_d)=0.
    \label{eq:int_tilde=0}
\end{align}
Then we define
\begin{align}
    F(x_1,\cdots,x_d)
    := \frac{1}{d}\sum_{i=1}^{d} \int_{0}^{2\pi}{\rm d}y_1 \cdots \int_{0}^{x_i}{\rm d}y_i \cdots \int_{0}^{2\pi}{\rm d}y_d\ \tilde{\tilde{f}}(y_1,\cdots,y_d)
\end{align}
$F$ is in $C_{\rm c}^\infty(T^d)$ because of (\ref{eq:int_tilde=0}).
We can see $\Delta^{\rm cl}F^\star=f$, thus $f\in {\rm im}(\Delta^{\rm cl})$ and ${\rm im}(\Delta^{\rm BV})\supset {\rm ker}(\pi^0)$.

Third, we show $H^{-1}(A)=H^{-1}(B)$. Clearly $H^{-1}(B)=\mathbb{C}$. Then let us think of
\begin{align}
    H^{-1}(A)={\rm ker}(\Delta^{\rm cl}).
\end{align}
This is $\mathbb{C}$, since the non-trivial solutions of 
\begin{align}
    \Delta f=0
\end{align}
are $f={\rm const}\in C_{\rm c}^\infty(T^d)$.
And the action of $\pi^{-1}$ is just a multiplication by $(2\pi)^d$. Then $H^{-1}(A)=H^{-1}(B)=\mathbb{C}$.
Thus $A$ and $B$ are quasi-isomorphic.
\end{proof}

Therefore, the inclusion map $i$ does NOT give a state.
\begin{align}
    i:
    H^0{\rm Obs}^{\rm cl}(U)
    \to
    H^0{\rm Obs}^{\rm cl}(T^d)
    =
    \mathbb{C}[q]
\end{align}
However, by sending the generator $q$ to $0$, we obtain a state.
\begin{tcolorbox}[colframe=red,colback=red!3!]
\begin{definition}\ \\
    A compactification state $\langle-\rangle_{\rm cptf}$ is defined as $j\circ i$ where $j$ sends $q$ to $0$.
    \begin{align}
    \langle-\rangle_{\rm cptf}:
    H^0{\rm Obs}^{\rm cl}(U)
    \xrightarrow{i}
    \mathbb{C}[q]
    \xrightarrow{j}
    \mathbb{C}
    \end{align}
    where $U$ is bounded.
\end{definition}
\end{tcolorbox}
\subsection{\texorpdfstring{The map $j$ and IR divergence}{The map j and IR divergence}}\label{subsec:why_j_?}
In order to see the physical meaning of the map $j$, 
we review the following theorem in a one-dimensional system.
\begin{tcolorbox}[colframe=blue,colback=blue!3!]
\begin{theorem}\label{thm:M=I_cl}
{\rm (K. Costello, O. Gwilliam\cite[\S 4.2.4]{Costello:2016vjw})}\\
    If $I\subset \mathbb{R}$ is an interval, 
    \begin{align}
        H^n\left({\rm Obs}^{\rm cl}(I)\right)
        =
        \left\{
        \begin{array}{ll}
        \mathbb{C}[q,p] & (n=0) \\
        0 & ({\rm otherwise})
        \end{array}
        \right.
    \end{align}
    where $q,p$ has degree-0.
\end{theorem}
\end{tcolorbox}
\begin{proof}
    We show a quasi-isomorphism:
    \begin{align}
        \pi:A\to B
    \end{align}
    where
    \begin{align}
        A=\left(C_{\rm c}^\infty(I)^{-1}\xrightarrow{\Delta^{\rm cl}}C_{\rm c}^\infty(I)^0\right),\ 
        B=(0\to\mathbb{C}^2).
    \end{align}
    $\mathbb{C}^2$ sits in degree-0. And we denote the bases of $\mathbb{C}^2$ as $\{q,p\}$.
    The cohomology of ${\rm Sym}(A)$ and ${\rm Sym}(B)$ are $H^*({\rm Obs}^{\rm cl}(I))$ and $\mathbb{C}[q,p]$, respectively.
    
    First of all, we show that the following diagram commutes:
    \begin{align}
        \xymatrix{
    C_{\rm c}^\infty(I)^{-1} \ar[r] \ar[d]_{\pi^{-1}} & C_{\rm c}^\infty(I)^0 \ar[d]_{\pi^0} \\
    0 \ar[r] & \mathbb{C}^2
  }\label{eq:commutative_diagram_in_R1}
    \end{align}
    where $\pi^0$ is defined as  
   \begin{align}
       \pi^0(g)
       :=q\int_I {\rm d}x\ g(x)\phi_q(x)+p\int_I {\rm d}x\ g(x)\phi_p(x)
   \end{align}
   for $g\in C_{\rm c}^\infty(I)^0$.
   The definition of $\phi_q,\phi_p\in C^\infty(\mathbb{R})$ is as follows.
    In the case where $m>0$ we define $\phi_q,\phi_p\in C^\infty(\mathbb{R})$ as
\begin{align}
    \phi_q(x)=\frac{1}{2}(e^{mx}+e^{-mx}),\ 
    \phi_p(x)=\frac{1}{2m}(e^{mx}-e^{-mx}).
\end{align}
They form the kernel of $-\Delta+m^2$.
If $m=0$, we define
\begin{align}
    \phi_q(x)=1,\ 
    \phi_p(x)=x.
\end{align}
We note that $\phi_p'(x)=\phi_q(x)$.
We can easily check that $\pi^0(g)=0$ holds if $g=\Delta^{\rm cl}f^{\star}$ for some $f^{\star} \in C_{\rm c}^\infty(I)^{-1}$.
Hence the diagram \eqref{eq:commutative_diagram_in_R1} commutes.

Next, we show $H^0(A)=H^0(B)$. By definition,
\begin{align}
    H^0(A)
    =
    \frac{C_{\rm c}^\infty(I)^0}
    {{\rm im}(\Delta^{\rm cl})}.
\end{align}
We show that ${\rm im}(\Delta^{\rm cl})={\rm ker}(\pi^0)$. If it holds, $H^0(A)=H^0(B)$. Obviously,
\begin{align}
    {\rm im}(\Delta^{\rm cl})\subset {\rm ker}(\pi^0).
\end{align}
Then we check ${\rm im}(\Delta^{\rm cl})\supset {\rm ker}(\pi^0)$.
Take $f\in {\rm ker}(\pi^0)$. $f$ satisfies
\begin{align}
    &\int_I {\rm d}x\ f(x)e^{mx}=0\ {\rm and}
    \int_I {\rm d}x\ f(x)e^{-mx}=0\ \ 
    ({\rm for\ massive\ case}),\notag\\
    &\int_I {\rm d}x\ f(x)x=0\ {\rm and}
    \int_I {\rm d}x\ f(x)=0\ \ 
    ({\rm for\ massless\ case}) \label{eq:integral_of_f},
\end{align}
because $\int_I {\rm d}x f(x)\phi_p(x)=0$ and $\int_I {\rm d}x f(x)\phi_q(x)=0$.
Let $G\in C^0(\mathbb{R})$ be the Green function:
\begin{align}
    &G(x)=\frac{1}{2m}e^{-m|x|}\ \ 
    ({\rm for\ massive\ case}),\notag\\
    &G(x)=-\frac{1}{2}|x|\ \ 
    ({\rm for\ massless\ case}).
\end{align}
Then the convolution of $f$ and $G$ is
\begin{align}
    (G\cdot f)(x):=\int_I {\rm d}y\ G(x-y)f(y).
\end{align}
This is in $C_{\rm c}^\infty(I)^0$ because of  (\ref{eq:integral_of_f}). For $(G\cdot f)^\star\in C_{\rm c}^\infty(I)^{-1}$, 
\begin{align}
    f=\Delta^{\rm cl}(G\cdot f)^\star.
\end{align}
Thus $f\in {\rm im}(\Delta^{\rm cl})$, and ${\rm im}(\Delta^{\rm cl})\supset {\rm ker}(\pi^0)$.

Finally, we show $H^{-1}(A)=H^{-1}(B)$. Clearly $H^{-1}(B)=0$. Then let us think of
\begin{align}
    H^{-1}(A)={\rm ker}(\Delta^{\rm cl}).
\end{align}
This is trivial since there is only a trivial solution for 
\begin{align}
    (-\Delta+m^2)f=0
\end{align}
for $f\in C_{\rm c}^\infty(I)$. Then $H^{-1}(A)=H^{-1}(B)=0$.
\end{proof}

In the massless case, the map $j$ should be interpreted as the operation that removes the infrared divergence.
This is because the generator $q$ corresponds to the constant field configuration $\phi_q(x)=1$ appearing in the definition of $\pi^0$, i.e. to a long-wavelength mode. (In contrast, $\phi_p(x)$ does not lie on $T^{d=1}$.)

In a massive theory, there are no long-wavelength modes, and hence the expectation values $\langle-\rangle_{\rm cptf}$ can be determined directly.
In contrast, a massless theory does contain long-wavelength modes, and these give rise to IR divergences.
In the path integral formulation, one must remove such modes in order to introduce well-defined expectation values.
Setting the generator $q$ to zero precisely implements this removal.

\section{Schwartz state}

\subsection{Massive case}

Although the definition of the compactification state is conceptually straightforward, it requires a locality condition in order to be well defined.
This condition is automatically satisfied for local observables, but extended observables may fail to obey it.
In contrast, the Schwartz state is well defined without imposing locality, which makes it applicable to gauge theories containing extended observables such as Wilson loops.

Functions with compact support can be regarded as Schwartz functions in the usual way.
Thus we obtain an inclusion
\begin{align}
    i:C_{\rm c}^\infty(\mathbb{R}^d)
    \hookrightarrow
    \mathcal{S}(\mathbb{R}^d)
\end{align}
where $\mathcal{S}(\mathbb{R}^d)$ denotes the Schwartz space.
We define
\begin{align}
    {\rm Obs}^{\rm cl}_{\mathcal{S}}(\mathbb{R}^d)
    :=
    {\rm Sym}
    \left(
    \mathcal{S}(\mathbb{R}^d)^{-1}
    \xrightarrow{\Delta^{\rm cl}}
    \mathcal{S}(\mathbb{R}^d)^{0}
    \right).
\end{align}
Then we have
\begin{align}
    i:
    {\rm Obs}^{\rm cl}(\mathbb{R}^d)
    \to
    {\rm Obs}^{\rm cl}_{\mathcal{S}}(\mathbb{R}^d)
\end{align}
and
\begin{align}
    i:
    H^*{\rm Obs}^{\rm cl}(\mathbb{R}^d)
    \to
    H^*{\rm Obs}^{\rm cl}_{\mathcal{S}}(\mathbb{R}^d).
\end{align}
In fact, the inclusion map $i$ defines a state. To see this, we introduce the following theorem.
\begin{tcolorbox}[colframe=blue,colback=blue!3!]
\begin{theorem}
{\rm (K. Costello, O. Gwilliam\cite[\S 4.9]{Costello:2016vjw})}\\
    In the massive case,
    \begin{align}
    H^n{\rm Obs}^{\rm cl}_{\mathcal{S}}(\mathbb{R}^d)
    =
    \left\{
    \begin{array}{ll}
    \mathbb{C} & (n=0) \\
    0 & ({\rm otherwise})
    \end{array}
    \right.
\end{align}
\end{theorem}
\end{tcolorbox}
\begin{proof}
    \begin{align}
        A:=
        \left(\mathcal{S}
        (\mathbb{R}^d)^{-1}
        \xrightarrow{\Delta^{\rm cl}}
        \mathcal{S}
        (\mathbb{R}^d)^0\right)
    \end{align}
    This is an isomorphism. To see this, we consider the Fourier transform:
    \begin{align}
        \begin{array}{rccc}
        &\mathcal{S}(\mathbb{R}^d)&\longrightarrow& \mathcal{S}(\mathbb{R}^d)\\
        & \rotatebox{90}{$\in$}&  & \rotatebox{90}{$\in$} \\
        & f   & \longmapsto   & \hat{f}
        \end{array}
    \end{align}
    where 
    \begin{align}
        \hat f(k_1,\cdots,k_d)
        :=
        \int_{\mathbb{R}^d} f(x_1,\cdots,x_d)e^{ik\cdot x}
        {\rm d}x.
    \end{align}
    The Fourier transform gives an isomorphism 
    $\mathcal{S}(\mathbb{R}^d)\stackrel{\sim}{\longrightarrow}\mathcal{S}(\mathbb{R}^d)$.
    Then we have
    \begin{align}
        \hat{A}
        :=
        \left(\mathcal{S}
        (\mathbb{R}^d)^{-1}
        \xrightarrow{\hat\Delta^{\rm cl}}
        \mathcal{S}
        (\mathbb{R}^d)^0\right)
    \end{align}
    where $(\hat\Delta^{\rm cl} \hat f)(k)=-(k^2+m^2)\hat{f}(k)$.
    Since it is massive, $\hat{\Delta}^{\rm cl}$ is an isomorphism. Then $\Delta^{\rm cl}$ is also an isomorphism.
    Thus $H^*(A)=0$, then $H^0({\rm Sym}(A))=\mathbb{C}$ and $H^{n\le -1}({\rm Sym}(A))=0$
\end{proof}
Then the inclusion map $i$ gives a state $\langle-\rangle_{\rm Sch}$.
\footnote{The abbreviation “${\rm Sch}$” stands for  ``Schwartz."}
\begin{tcolorbox}[colframe=red,colback=red!3!]
\begin{definition}
{\rm (K. Costello, O. Gwilliam\cite{Costello:2016vjw})}\\
    The Schwartz state $\langle-\rangle_{\rm Sch}$ is the smooth map:
    \begin{align}
    \langle-\rangle_{\rm Sch}:
        H^0{\rm Obs}^{\rm cl}(\mathbb{R}^d)
    \to
    H^0{\rm Obs}^{\rm cl}_{\mathcal{S}}(\mathbb{R}^d)
    =\mathbb{C}
    \end{align}
\end{definition}
\end{tcolorbox}

\subsection{Massless case in one-dimension}
In the massless case, $H^0{\rm Obs}^{\rm cl}_\mathcal{S}(\mathbb{R}^d)$ is NOT isomorphic to $\mathbb{C}$.
This is because
\begin{align}
    \hat\Delta^{\rm cl}:
    \mathcal{S}(\mathbb{R}^d)\ni
    \hat{f}(k)\mapsto -k^2 \hat{f}(k)
    \in\mathcal{S}(\mathbb{R}^d)
\end{align}
is not an isomorphism since 
\begin{align}
    ({\hat\Delta^{\rm cl}})^{-1}:
    \hat{f}(k)\mapsto -\frac{\hat{f}(k)}{k^2}
\end{align}
is not well-defined.
Physicists call it the IR divergence.

However, the IR divergence can be controlled by the following theorem.
Using this, we can define the Schwartz state in the massless case.
\begin{tcolorbox}[colframe=blue,colback=blue!3!]
\begin{theorem}\label{thm:IR_one-dim}\ \\
    Let $\hat f\in \mathcal{S}(\mathbb{R})$.
    If $\hat f$ satisfies
    \begin{align}
        \hat f(0)=0,\\
        \partial \hat{f}(0)=0,
    \end{align}
    then there exists a Schwartz function $\hat{h}$ on $\mathbb{R}$ such that
    \begin{align}
        \hat{f} (k) = -k^2 \hat{h}(k).
    \end{align}
\end{theorem}
\end{tcolorbox}
\begin{proof}
    First of all, we find $\hat g\in\mathcal{S}(\mathbb{R})$ satisfying
    \begin{align}
        \hat{f}(k)=
        k\hat{g}(k).
    \end{align}
    Since $\hat f(0)=0$, then
    \begin{align}
        \hat f(k)
        =
        \hat f(k)
        -
        \hat f(0).
    \end{align}
    We have a line integral
    \begin{align}
        \hat f(k)
        -
        \hat f(0)
        &=
        \int_0^k 
        {\rm d}p\ 
        (\partial\hat{f})(p)
        \notag\\
        &=
        k
        \int_0^1 
        {\rm d}t\ 
        (\partial\hat{f})(kt).
    \end{align}
    where $t:=p/k$.
    Define 
    \begin{align}
    \hat g(k)
    :=  \int_0^1
    {\rm d}t\ 
         (\partial \hat f)(kt),
    \end{align}
    then
    \begin{align}
        \hat{f}(k)
        =
        k\hat{g}(k)\label{eq:f=kg}.
    \end{align}
    $\hat{g}$ is Schwartz, since we can interchange differentiation and integration, because $\partial \hat{f}$ is Schwartz.

    Now $\hat{g}(0)=0$ by $\partial \hat f(0)=0$. By the same procedure as the above one, we have $\hat{h}\in\mathcal{S}(\mathbb{R})$ satisfying
    \begin{align}
        \hat{g}(k)=k\hat{h}(k)\label{eq:g=kh}.
    \end{align}

    By (\ref{eq:f=kg}) and (\ref{eq:g=kh}),
    \begin{align}
        \hat{f}(k)=k^2\hat{h}(k).
    \end{align}
\end{proof}

By Theorem \ref{thm:IR_one-dim}, we obtain the following theorem.
\begin{tcolorbox}[colframe=blue,colback=blue!3!]
\begin{theorem}\label{thm:Schwatz_massless}\ \\
    In the massless and $d=1$ case 
    \begin{align}
        H^n{\rm Obs}^{\rm cl}_{\mathcal{S}}(\mathbb{R})
        =
        \left\{
        \begin{array}{ll}
        \mathbb{C}[q,p] & (n=0) \\
        0 & ({\rm otherwise})
        \end{array}
        \right.
    \end{align}
\end{theorem}
\end{tcolorbox}

\begin{proof}
We show a quasi-isomorphism:
\begin{align}
    \pi:A\to B
\end{align}
where
\begin{align}
    A=
    \left(\mathcal{S}(\mathbb{R})^{-1}\xrightarrow{\Delta^{\rm cl}}\mathcal{S}(\mathbb{R})^0\right),\ 
    B=(0\to\mathbb{C}^2).
\end{align}
$\mathbb{C}^2$ sits in degree-0. And we denote the bases of $\mathbb{C}^2$ as $\{q,p\}$.
The cohomology of ${\rm Sym}(A)$ and ${\rm Sym}(B)$ are $H^*{\rm Obs}^{\rm cl}_{\mathcal{S}}(\mathbb{R})$ and $\mathbb{C}[q,p]$ respectively.
    
First of all, we show the following commutative diagram:
\begin{align}
    \xymatrix{
\mathcal{S}(\mathbb{R})^{-1} \ar[r] \ar[d]_{\pi^{-1}} & \mathcal{S}(\mathbb{R})^0 \ar[d]_{\pi^0} \\
0 \ar[r] & \mathbb{C}^2
    }
\end{align}
where $\pi^0$ is defined as  
\begin{align}
   \pi^0(g)
   :=
   q\int_{\mathbb{R}} {\rm d}x
   \ g(x)\phi_q(x)
   +
   p\int_{\mathbb{R}} {\rm d}x
   \ g(x)\phi_{p_1}(x)
   \label{eq:pi^0}
\end{align}
for $g\in \mathcal{S}(\mathbb{R})^0$.
The definition of $\phi_q,\phi_{p}\in C^\infty(\mathbb{R})$ is as follows.
\begin{align}
\begin{cases}
    \phi_q(x)=1, \\
    \phi_{p}(x)=x.
  \end{cases}
\end{align}
They form the kernel of $-\Delta^{\rm cl}$.
We can easily check that
   $\pi^0(g)=0$ holds if $g=\Delta^{\rm cl}f^{\star}.$
Note that $\pi^0$ is well-defined, in other words all integrations in (\ref{eq:pi^0}) converge, since products of Schwartz function $g$ and polynomials $\phi_q,~\phi_{p}$ are also Schwartz functions.

Next, we show $H^0(A)=H^0(B)$. By definition,
\begin{align}
    H^0(A)
    =
    \frac{\mathcal{S}(\mathbb{R})^0}
    {{\rm im}(\Delta^{\rm cl})}.
\end{align}
$\pi^0$ is surjective \footnote{The proof is in Appendix \ref{subsec:pi^0_surjective}.},
then by the isomorphism theorem
\begin{align}
    H^0(B)
    =
    \frac{\mathcal{S}(\mathbb{R})^0}
    {{\rm ker}(\pi^0)}.
\end{align}
We show that ${\rm im}(\Delta^{\rm cl})={\rm ker}(\pi^0)$. If it holds, $H^0(A)=H^0(B)$. Obviously,
\begin{align}
    {\rm im}(\Delta^{\rm cl})\subset {\rm ker}(\pi^0).
\end{align}
Then we check ${\rm im}(\Delta^{\rm cl})\supset {\rm ker}(\pi^0)$.
Take $f\in {\rm ker}(\pi^0)$. $f$ satisfies
\begin{align}
    &\int_{\mathbb{R}}f(x)\ 
    {\rm d}x
    =0,\label{eq:pi^0_q}\\
    &\int_{\mathbb{R}}xf(x)\ 
    {\rm d}x
    =0\label{eq:pi^0_p}.
\end{align}
In order to make it clear, we consider the Fourier transform of $A$.
\begin{align}
    \hat A
    =
    \left(\mathcal{S}(\mathbb{R})^{-1}\xrightarrow{\hat\Delta^{\rm cl}}\mathcal{S}(\mathbb{R})^0\right)
\end{align}
where $\hat\Delta^{\rm cl}:\hat f(k)\mapsto -k^2f(k)$. 
And the above conditions (\ref{eq:pi^0_p}) (\ref{eq:pi^0_q}) are the same as
\begin{align}
   \hat f(0)=0,\\
   \partial\hat f(0)=0.
\end{align}
Under these conditions 
\begin{align}
    \hat h(k):=-\frac{\hat{f}(k)}{k^2}
\end{align}
is in $\mathcal{S}(\mathbb{R})$ by Theorem \ref{thm:IR_one-dim}.
Hence, $\hat \Delta^{\rm cl} \hat h=\hat f$ or $\Delta^{\rm cl} h=f$. 
Thus $f\in {\rm im}(\Delta^{\rm cl})$, and ${\rm im}(\Delta^{\rm cl})\supset {\rm ker}(\pi^0)$.

Finally, we show $H^{-1}(A)=H^{-1}(B)$. Clearly $H^{-1}(B)=0$. Then let us think of
\begin{align}
    H^{-1}(A)={\rm ker}(\Delta^{\rm cl}).
\end{align}
This is trivial since the solution of 
\begin{align}
    \Delta f(x)=0
    \ 
    {\rm or}
    \ 
    -k^2\hat f(k)=0
\end{align}
is $f=0$ or $\hat f=0$.
Then $H^{-1}(A)=H^{-1}(B)=0$.
\end{proof}

We now obtain 
\begin{align}
H^n{\rm Obs}^{\rm cl}_{\mathcal{S}}(\mathbb{R}) =\mathbb{C}[q,p].
\end{align}
This allows us to define the Schwartz state also in the massless case.\footnote{Costello and Gwilliam define the Schwartz state only for massive theories; the massless case is not treated in their work.}

\begin{tcolorbox}[colframe=red,colback=red!3!]
\begin{definition}\ \\
    The Schwartz state $\langle-\rangle_{\rm Sch}$ is the smooth map:
    \begin{align}
    \langle-\rangle_{\rm Sch}:
        H^0{\rm Obs}^{\rm cl}(\mathbb{R})
    \to
    H^0{\rm Obs}^{\rm cl}_{\mathcal{S}}(\mathbb{R})
    =\mathbb{C}[q,p]
    \xrightarrow{j} \mathbb{C}
    \end{align}
    where $j$ is a map: $q,p\mapsto 0$.
\end{definition}
\end{tcolorbox}

In the case of the compactification state, sending the generator $q\mapsto 0$ means to get rid of IR divergences as explained in \ref{subsec:why_j_?}.
However, in the case of the Schwartz state, the map $j$ sends not only $q$ but also $p$ to $0$.
In terms of standard physical terminology, the map $p\mapsto 0$ corresponds to the assumption that the state $|0\rangle$ satisfies
\begin{align}
    P|0\rangle=0.
\end{align}
In other words, it expresses translation invariance. 

The compactification state automatically has translation invariance because we impose periodic boundary conditions. In contrast, translation invariance must be assumed for the Schwartz state.

\subsection{Massless case in higher dimension }
In the one-dimensional case, we show
\begin{align}
        \left(\mathcal{S}(\mathbb{R})^{-1}\xrightarrow{\Delta^{\rm cl}}\mathcal{S}(\mathbb{R})^0\right)
        \xrightarrow[\ \pi\ ]{\sim}{}
        (0\to\mathbb{C}^2).
\end{align}
The $\mathbb{C}^2$ is the solution space of the Laplace equation $\Delta \phi=0$ in one-dimensional space.

In order to generalize to the higher dimensional case, we need to consider the following solution space\footnote{$\mathscr{H}$ denotes the space of harmonics.}:
\begin{align}
    \mathscr{H}
    :=
    \{\phi\in \mathcal{S}'(\mathbb{R}^d)\ |\ \Delta_{\mathcal{S}'}\phi=0\}
\end{align}
where $\mathcal{S}'(\mathbb{R}^d)$ is the space of tempered distributions and $\Delta_{\mathcal{S}'}$ is the Laplacian in $\mathcal{S}'(\mathbb{R}^d)$.
$\mathcal{S}'(\mathbb{R}^d)$ is naturally the dual space of $\mathcal{S}(\mathbb{R}^d)$, and therefore the product $\langle \phi,f\rangle$ is well-defined for $f\in \mathcal{S}(\mathbb{R}^d)$ and $\phi\in\mathcal{S}'(\mathbb{R}^d)$.
If $\phi$ is a function, we can represent $\langle \phi,f\rangle$ as integration:
\begin{align}
    \langle \phi,f\rangle
    =
    \int_{\mathbb{R}^d} {\rm d}x_1\cdots{\rm d}x_d
    \ 
    \overline{\phi(x_1,\cdots,x_d)}\ f(x_1,\cdots,x_d).
\end{align}

Later, we show a quasi-isomorphism: 
\begin{align}
        \left(\mathcal{S}(\mathbb{R}^d)^{-1}\xrightarrow{\Delta^{\rm cl}}\mathcal{S}(\mathbb{R}^d)^0\right)
        \xrightarrow[\ \pi\ ]{\sim}{}
        (0\to\mathscr{H}).
\end{align}
Roughly speaking, $\pi$ is defined\footnote{This is not well-defined because $\mathscr{H}$ is infinite-dimensional. More precisely, we need to consider the completion of $\mathscr{H}$ and modify the definition of $\pi$.} as 
\begin{align}
    \begin{array}{rccc}
        \pi^0\colon
        &\mathcal{S}(\mathbb{R}^d)^0&\longrightarrow& \mathscr{H}\\
        & \rotatebox{90}{$\in$}&  & \rotatebox{90}{$\in$} \\
        & f   & \longmapsto   & 
        \sum_{\phi\in B(\mathscr{H})}\phi\langle \phi,f\rangle.
        \end{array}
\end{align}
where $B(\mathscr{H})$ is a basis of $\mathscr{H}$.
This is a natural generalization of the one-dimensional case.

In order to make clear the meaning of $B(\mathscr{H})$, 
we introduce the following known theorem.
\begin{tcolorbox}[colframe=blue,colback=blue!3!]
\begin{theorem}\ \\
    Let $\phi\in \mathcal{S}'(\mathbb{R}^d)$.
    If $\Delta_{\mathcal{S'}}\phi=0$, then $\phi$ can be represented as a polynomial.
\end{theorem}
\end{tcolorbox}
\noindent{} In other words, $\phi\in\mathcal{H}$ is a harmonic polynomial.
For polynomials, the Fisher inner product is convenient.
\begin{tcolorbox}[colframe=red,colback=red!3!]
\begin{definition}{\rm (The Fisher inner product)} \\
Let $\phi,\psi$ be polynomials on $\mathbb{R}^d$.
Fisher inner product is defined as
\begin{align}
    (\phi|\psi)
    :=\left[
    \overline{\phi\left(\frac{\partial}{\partial x_1},
    \cdots,
    \frac{\partial}{\partial x_1}\right)}
    \psi(x_1,\cdots,x_d)\right]_{x_1=\cdots x_d=0}.
\end{align}
\end{definition}
\end{tcolorbox}
\noindent{}Let $\phi$ be $k$-degree  and $\psi$ be $l$-degree.
If $k\neq l$ then $(\phi|\psi)=0$.
Thus we obtain an orthogonal decomposition of $\mathscr{H}$.
\begin{align}
    \mathscr{H}
    =
    \bigoplus_{k=0}^\infty \mathscr{H}_k.
\end{align}
Here $\mathscr{H}_k$ denotes the subspace of $k$-degree polynomials.
The dimension of $\mathscr{H}_k$ is finite and given by
\begin{align}
    {\rm dim}(\mathscr{H}_k)
    =
    \frac{(d+2k-2)(d+k-3)!}{(d-2)!k!}.
\end{align}
We apply the Gram–Schmidt process to obtain the basis $B(\mathscr{H})=\bigcup_{k=0}^\infty B(\mathscr{H}_k)$.

\begin{tcolorbox}[colframe=blue,colback=blue!3!]
\begin{theorem}\label{thm:higher_dim}\ \\
    In the massless and higher dimensional case
    \begin{align}
    H^n{\rm Obs}^{\rm cl}_{\mathcal{S}}(\mathbb{R}^d)
    =
    \left\{
    \begin{array}{ll}
    {\rm Sym}(\widetilde{\mathscr{H}}) & (n=0) \\
    0 & ({\rm otherwise})
    \end{array}
    \right.
    \end{align}
    where $\widetilde{\mathscr{H}}$ is a subspace of the direct product $\prod_k \mathscr{H}_k$ and the cardinality of its bases are at least countably infinite.
\end{theorem}
\end{tcolorbox}
The proof is rather long, so we provide the details in Appendix~\ref{sec:proof_of_thorem_5_3_2}.

\begin{tcolorbox}[colframe=red,colback=red!3!]
\begin{definition}\ \\
    The Schwartz state $\langle-\rangle_{\rm Sch}$ is the smooth map:
    \begin{align}
    \langle-\rangle_{\rm Sch}:
        H^0{\rm Obs}^{\rm cl}(\mathbb{R}^d)
    \to
    H^0{\rm Obs}^{\rm cl}_{\mathcal{S}}(\mathbb{R}^d)
    ={\rm Sym}(\widetilde{\mathscr{H}})
    \xrightarrow{j} \mathbb{C}
    \end{align}
    where $j$ sends all generators to $0$.
\end{definition}
\end{tcolorbox}
In the one-dimensional case, the condition $p,q\mapsto 0$
simply reflected a single requirement: translation invariance,
\begin{align}
    P|0\rangle=0.
\end{align}
However, in higher dimensions the situation is more complicated, since there are infinitely many generators rather than just two.
This suggests that additional invariance conditions may be required.

\section{Equivalence}
\subsection{Massive case}
As mentioned at the beginning, we show the equivalence of the natural augmentation state, the compactification state, and the Schwartz state in the massive case.
First of all, we introduce the following theorem shown by Costello and Gwilliam\cite{Costello:2016vjw}.
\begin{tcolorbox}[colframe=blue,colback=blue!3!]
\begin{theorem}
{\rm (K. Costello, O. Gwilliam\cite[\S 4.9]{Costello:2016vjw})}
\label{thm:aug=Sch}\\
    In the massive case,
    \begin{align}
    \langle-\rangle_{\rm aug}
    =
    \langle-\rangle_{\rm Sch}.
\end{align}
\end{theorem}
\end{tcolorbox}
\noindent{}In addition, we show that $\langle-\rangle_{\rm aug}$ is the same as $\langle-\rangle_{\rm cptf}$ in the massive cases.
\begin{tcolorbox}[colframe=blue,colback=blue!3!]
\begin{theorem}\ \\
    In the massive case, for observables that satisfy the locality condition, we have
    \begin{align}
    \langle-\rangle_{\rm aug}
    =
    \langle-\rangle_{\rm cptf}.
\end{align}
\end{theorem}
\end{tcolorbox}
\begin{proof}
Take $\mathcal{O}\in {\rm Obs}^{\rm cl}(U)\subset {\rm Obs}^{\rm cl}(T^d)$, and calculate $\langle\mathcal{O}\rangle_{\rm cptf}$. 
\begin{align}
    \Delta^{\rm cl}_{T^d}
    :
    C_{\rm c}^\infty(T^d)^{-1}
    \to
    C_{\rm c}^\infty(T^d)^0
\end{align}
is an isomorphism. Therefore
\begin{align}
    \mathcal{O}
    =
    c+f+f_1*f_2+\cdots
    =
    c+\Delta^{\rm cl}_{T^d}(\cdots).
\end{align}
Hence,
\begin{align}
    \langle\mathcal{O}\rangle_{\rm cptf}=c=\langle\mathcal{O}\rangle_{\rm aug}.
\end{align}

\end{proof}
Combining the two theorems above, we obtain the equivalence of the three states in the massive case.

\subsection{Massless case}
We show the equivalence of  the three states in the massless case.
\begin{tcolorbox}[colframe=blue,colback=blue!3!]
\begin{theorem}\ \\
    In the massless case, for observables that satisfy the locality condition, we have
    \begin{align}
    \langle-\rangle_{\rm aug}
    =
    \langle-\rangle_{\rm cptf}.
\end{align}
\end{theorem}
\end{tcolorbox}
\begin{proof}
Take $\mathcal{O}\in {\rm Obs}^{\rm cl}(U)$, and calculate $\langle\mathcal{O}\rangle_{\rm cptf}$.
\begin{align}
    \langle\mathcal{O}\rangle_{\rm cptf}
    =j\circ\pi^0(\mathcal{O})
\end{align}
$\mathcal{O}=c+f+f_1*f_2+\cdots$, then
\begin{align}
    \pi^0(c+f+f_1*f_2+\cdots)
    &=c+\pi^0(f)+\pi^0(f_1)\cdot \pi^0(f_2)+\cdots\notag\\
    &=c+q\times ({\rm some\ number}).
\end{align}
Hence,
\begin{align}
    \langle\mathcal{O}\rangle_{\rm cptf}=c=\langle\mathcal{O}\rangle_{\rm aug}.
\end{align}
\end{proof}

\begin{tcolorbox}[colframe=blue,colback=blue!3!]
\begin{theorem}\ \\
    In the massless case,
    \begin{align}
    \langle-\rangle_{\rm aug}
    =
    \langle-\rangle_{\rm Sch}.
\end{align}
\end{theorem}
\end{tcolorbox}
\begin{proof}
Take $\mathcal{O}\in {\rm Obs}^{\rm cl}(\mathbb{R}^d)$, and calculate $\langle\mathcal{O}\rangle_{\rm Sch}$.
\begin{align}
    \langle\mathcal{O}\rangle_{\rm Sch}
    =j\circ\pi^0(\mathcal{O})
\end{align}
$\mathcal{O}=c+f+f_1*f_2+\cdots$, then
\begin{align}
    \pi^0(c+f+f_1*f_2+\cdots)
    &=c+\pi^0(f)+\pi^0(f_1)\cdot \pi^0(f_2)+\cdots\notag\\
    &=c+({\rm some\ generator})\times ({\rm some\ number}).
\end{align}
Hence,
\begin{align}
    \langle\mathcal{O}\rangle_{\rm Sch}=c=\langle\mathcal{O}\rangle_{\rm aug}.
\end{align}
\end{proof}
Combining the two theorems above, we obtain the equivalence of the three states also in the massless case.

\section{Discussion}

\subsection{One-dimensional case vs. higher-dimensional cases}

We have seen the concrete constructions of states in factorization algebras. 
In the case of the compactification state,
\begin{align}
    \langle-\rangle_{\rm cptf}:
    {\rm Obs}^{\rm cl}(U)
    \to{\rm Obs}^{\rm cl}(T^d)
    \cong
  \begin{cases}
    \mathbb{C}     & ({\rm massive}) \\   
    \mathbb{C}[q]   \xrightarrow{j}\mathbb{C}  &({\rm massless}) .
  \end{cases}
\end{align}
Here the origin of the generator $q$ is the long-wavelength mode of the scalar field, so the map $j:q\mapsto 0$ means to get rid of IR divergences.
On the other hand, in the case of the Schwartz state,
\begin{align}
    \langle-\rangle_{\rm Sch}:
    {\rm Obs}^{\rm cl}(\mathbb{R}^d)
    \to{\rm Obs}^{\rm cl}_{\mathcal{S}}(\mathbb{R}^d)
    \cong
  \begin{cases}
    \mathbb{C}     & ({\rm massive}) \\   
    {\rm Sym}(\widetilde{\mathscr{H}})   \xrightarrow{j}\mathbb{C}  &({\rm massless}) .
  \end{cases}
\end{align}
Especially, in the one-dimensional case, ${\rm Sym}(\widetilde{\mathscr{H}}) $ is just $\mathbb{C}[q,p]$.
This space may reflect the symmetry present in the asymptotic region. For example, in the one-dimensional case, the map $j:p\mapsto 0$ means the assumption of the translation invariance for the state: $P|0\rangle=0$.

The dimension of ${\rm Sym}(\widetilde{\mathscr{H}})$ might be uncountably infinite. It seems to come from the difference between the one-dimensional case and the higher case.
In the former case, the commutation relation is
\begin{align}
    [q,p]=i,
\end{align}
while in the latter case, 
\begin{align}
    [\Phi(\bm{x}),\Pi(\bm{y})]=i\delta(\bm{x}-\bm{y}).
\end{align}
Thus we have the uncountable label $\bm{x}\in\mathbb{R}^{d-1}$ of $\Phi(x)$.

\subsection{Classical vs. quantum}

In this paper, we have discussed only the classical situation.
In this section, we examine how the story changes after quantization.

For simplicity, we restrict ourselves to the massive case.
In the quantum setting, the natural augmentation state and the Schwartz state can be defined in the same way as in the classical case, and it is known that they are equivalent, as shown in\cite[\S 4.9]{Costello:2016vjw}.

However, the compactification state is not equivalent to the other states as it stands.
In fact, the expectation value $\langle\mathcal{O}\rangle_{\rm cptf}$ depends on the radius $R$ of the torus $T^d$.
This is a feature that does not appear in the classical case.
Interestingly, the equivalence is restored only in the limit $R\to\infty$:
\begin{align}
\lim_{R\to\infty}
\langle\mathcal{O}\rangle_{\rm cptf}
=
\langle\mathcal{O}\rangle_{\rm aug}
=
\langle\mathcal{O}\rangle_{\rm Sch}.
\end{align}
This result is physically reasonable, since in the limit $R\to\infty$ the torus $T^d$ approaches $\mathbb{R}^d$.
This corresponds to describing the thermodynamic limit in the language of factorization algebras.
It would be an interesting future direction to investigate how this picture changes in the massless case.

\subsection{Future directions}

It is also interesting to investigate ${\rm Sym}(\widetilde{\mathscr{H}})$ in the two-dimensional case.
Generally, in order to avoid the Coleman-Mermin-Wagner theorem\cite{Coleman:1973ci,Mermin:1966fe}, we need conformal symmetry (the Virasoro algebra):
\begin{align}
    L_m|0\rangle=0.
\end{align}
We expect that the Virasoro algebra can be derived from ${\rm Sym}(\widetilde{\mathscr{H}})$.

Another interesting direction is the study of Koszul duality in massless theories.
In such cases, constructing the Koszul dual of the observable algebra is not straightforward due to a moduli space of vacua\footnote{
This point is mentioned in Section 8 of \cite{Paquette:2021cij}.
}.
We expect that our results provide an appropriate choice of vacuum, thereby resolving this issue.

\section*{Acknowledgment}
We would like to thank Yuma Furuta for helpful discussions.
The work of M. K. is supported by Grant-in-Aid for JSPS Fellows No. 22KJ1989.
The work of T. S. was supported by JST SPRING, Grant Number JPMJSP2110.

\appendix
\numberwithin{equation}{subsection}
\section{About the proof of Theorem \ref{thm:Schwatz_massless}}
\subsection{\texorpdfstring{The map $\pi^0$ is surjective}{π0 is surjective}}\label{subsec:pi^0_surjective}
\begin{tcolorbox}[colframe=blue,colback=blue!3!]
\begin{theorem}\ \\
    $\pi^0:\mathcal{S}(\mathbb{R})\to\mathbb{R}^2$ is surjective where
    $\pi^0$ is defined  as
    \begin{align}
        \pi^0(g)
        =
        q\int_{\mathbb{R}} {\rm d}x
       \ g(x)\phi_q(x)
       +
       p\int_{\mathbb{R}} {\rm d}x
       \ g(x)\phi_{p}(x)
    \end{align}
    and $q,p$ are a basis of $\mathbb{R}^2$.
\end{theorem}
\end{tcolorbox}
\begin{proof}
    It is enough to show that there are $Q,P\in\mathcal{S}(\mathbb{R})$ satisfying
    \begin{align}
        \pi^0(Q)=q,\ \pi^0(P)=p.
    \end{align}

    One good choice of $Q$ is a {\it smeared $\delta$-function}.
    \begin{align}
        Q(x)
        :=
        \delta_{\rm smeared}(x)
    \end{align}
    where we assume that $\delta_{\rm smeared}$ is even, has a compact support and satisfies
    \begin{align}
        \int{\rm d}x
        \ \delta_{\rm smeared}(x)=1.
    \end{align}
    $\phi_{p}$ is odd, then we have $\pi^0(\delta_{\rm smeared})=q$.
    And $P_i$ is given as
    \begin{align}
        P(x)
        :=
        -
        \frac{\partial}{\partial x}
        \delta_{\rm smeared}(x).
    \end{align}
    We can easily show that $\pi^0(P)=p$.

    Another choice of $Q$ is given by
    \begin{align}
        Q(x)
        :=
        \frac{1}{\sqrt{\pi}}\exp(x).
    \end{align}
    And $P$ is given as
    \begin{align}
        P(x)
        :=
        -\frac{\partial}{\partial x}Q(x)
        =
        \frac{2}{\sqrt{\pi}}x\exp(x^2).
    \end{align}
\end{proof}
\subsection{\texorpdfstring{Physical meanings of $q$ and $p$}{Physical meanings of q and p}}
By Theorem \ref{thm:Schwatz_massless}, we have an isomorphism:
\begin{align}
\begin{array}{rccc}
\pi^0\colon &H^0{\rm Obs}^{\rm cl}_\mathcal{S}(\mathbb{R}^d) &\stackrel{\sim}{\longrightarrow}& \mathbb{R}[q,p]\\
& \rotatebox{90}{$\in$}&  & \rotatebox{90}{$\in$} \\
& [Q]_\mathcal{S}      & \longmapsto   & q\\
& [P]_\mathcal{S}    & \longmapsto   & p.
\end{array}
\end{align}
Especially if we take $Q=\delta_{\rm smeared}$ and $P=-\partial\delta_{\rm smeared}$, $Q$ and $P$ are in $C_{\rm c}^\infty(\mathbb{R})$. Then we have an inclusion map:
\begin{align}
\begin{array}{rccc}
i\colon&H^0{\rm Obs}^{\rm cl}(\mathbb{R}) &\hookrightarrow& H^0{\rm Obs}^{\rm cl}_\mathcal{S}(\mathbb{R}^d)\\
& \rotatebox{90}{$\in$}&  & \rotatebox{90}{$\in$} \\
& [Q]   & \longmapsto   & [Q]_\mathcal{S}\\
& [P]    & \longmapsto  & [P]_\mathcal{S}.
\end{array}
\end{align}
Therefore $q$ and $p$ originate from the observables
\begin{align}
    Q=\delta_{\rm smeared},\ P=-\partial\delta_{\rm smeared}.
\end{align}
The action for the field $\Phi\in C^\infty(\mathbb{R}^d)$ is
\begin{align}
    &Q(\Phi)
    =\int_{\mathbb{R}} 
    {\rm d}x\ 
    \delta_{\rm smeared}(x)
    \Phi(x)
    \sim \Phi(0),\\ 
    &P(\Phi)
    =\int_{\mathbb{R}} 
    {\rm d}x\ 
    (-\partial\delta_{\rm smeared}(x))
    \Phi(x)
    \sim \partial\Phi(0).
\end{align}
These are the same as the position and momentum observables in physics literature.

\section{Some properties of harmonic polynomials}
\subsection{Hecke identities and a convenient representation of Fisher inner product}
\begin{tcolorbox}[colframe=blue,colback=blue!3!]
\begin{theorem}
{\rm \cite[Theorem 3.10]{stein1971introduction}}
\\
    If $\phi\in\mathscr{H}$, then we have two identities.
    \begin{align}
     \frac{1}{(2\pi)^{d/2}}
     \int_{\mathbb{R}^d}
     {\rm d}x_1\cdots{\rm d}x_d\ 
     \phi(x_1,\cdots,x_d)
     e^{-\frac{1}{2}x^2}
     e^{ik\cdot x}
     =
     \phi(ik_1,\cdots,ik_d)
     e^{-\frac{1}{2}k^2},\\
     \phi\left(
     \frac{\partial}{\partial x_1},
     \cdots,
     \frac{\partial}{\partial x_d}\right)
     e^{-\frac{1}{2}x^2}
     =
     \phi(-x_1,\cdots,-x_d)e^{-\frac{1}{2}x^2}.
    \end{align}
    We call them Hecke identities.
\end{theorem}
\end{tcolorbox}
\noindent{}By using Hecke identities, we have the following theorem.
\begin{tcolorbox}[colframe=blue,colback=blue!3!]
\begin{theorem}\ \\
    If $\phi,\psi\in\mathscr{H}$, then
    \begin{align}
     (\phi|\psi)
     &=
     \frac{1}{(2\pi)^{d/2}}
     \int_{\mathbb{R}^d}
     {\rm d}x_1\cdots{\rm d}x_d\ 
     \overline{\phi(x_1,\cdots,x_d)}
     \psi(x_1,\cdots,x_d)
     e^{-\frac{1}{2}x^2}\notag\\
     &=\left\langle \phi,\frac{1}{(2\pi)^{d/2}}\psi e^{-\frac{1}{2}x^2}\right\rangle
    \end{align}
\end{theorem}
\end{tcolorbox}
\begin{proof}
    $\overline{\phi\left(
     \frac{\partial}{\partial x_1},
     \cdots,
     \frac{\partial}{\partial x_d}\right)}
     \psi\left(x_1,\cdots,x_d\right)$ is also in $\mathscr{H}$,
    then by the first Hecke identities, we obtain
    \begin{align}
      \frac{1}{(2\pi)^{d/2}}
     \int_{\mathbb{R}^d}
     {\rm d}x_1\cdots{\rm d}x_d\ 
     \overline{\phi\left(
     \frac{\partial}{\partial x_1},
     \cdots,
     \frac{\partial}{\partial x_d}\right)}
     \psi\left(x_1,\cdots,x_d\right)
     e^{-\frac{1}{2}x^2}
     e^{ik\cdot x}\notag\\
     =
     \overline{\phi\left(-i
     \frac{\partial}{\partial k_1},
     \cdots,
     -i\frac{\partial}{\partial k_d}\right)}
     \psi\left(k_1,\cdots,k_d\right)
     e^{-\frac{1}{2}k^2}.
    \end{align}
    Substitute $k_1=\cdots=k_d=0$, then
    \begin{align}
     \frac{1}{(2\pi)^{d/2}}
     \int_{\mathbb{R}^d}
     {\rm d}x_1\cdots{\rm d}x_d\ 
     \overline{\phi\left(
     \frac{\partial}{\partial x_1},
     \cdots,
     \frac{\partial}{\partial x_d}\right)}
     \psi\left(x_1,\cdots,x_d\right)
     e^{-\frac{1}{2}x^2}
     =
     (\phi|\psi).
    \end{align}
    Integrate by part, 
    \begin{align}
    (\phi|\psi)
    &=
        \frac{1}{(2\pi)^{d/2}}
     \int_{\mathbb{R}^d}
     {\rm d}x_1\cdots{\rm d}x_d\ 
     \psi\left(x_1,\cdots,x_d\right)
     \overline{\phi\left(
     \frac{\partial}{\partial x_1},
     \cdots,
     \frac{\partial}{\partial x_d}\right)}
     e^{-\frac{1}{2}x^2}\notag\\
     &=
     \frac{1}{(2\pi)^{d/2}}
     \int_{\mathbb{R}^d}
     {\rm d}x_1\cdots{\rm d}x_d\ 
     \psi(x_1,\cdots,x_d)
     \overline{\phi(x_1,\cdots,x_d)}
     e^{-\frac{1}{2}x^2}.
    \end{align}
    We used the second Hecke identity in the last line.
\end{proof}

\section{About the proof of Theorem \ref{thm:higher_dim}}
\label{sec:proof_of_thorem_5_3_2}
\subsection{Outline of the proof}
The outline of the proof is the same as in Theorem \ref{thm:Schwatz_massless}.
We consider the following diagram
\begin{align}
    \xymatrix{
    \mathcal{S}(\mathbb{R}^d)^{-1} \ar[r] \ar[d]_{\pi^{-1}} & \mathcal{S}(\mathbb{R}^d)^0 \ar[d]_{\pi^0} \\
    0 \ar[r] & \widetilde{\mathscr{H}}
    }.
\end{align}

It seems to be enough to see the following things.
\begin{itemize}
    \item[(1)] The accurate definition of $\pi^0$ and $\widetilde{\mathscr{H}}$
    \item[(2)] The cardinality of the bases of $\widetilde{\mathscr{H}}$ is at least countably infinite.
    \item[(3)] ${\rm ker}(\pi^0)\cong{\rm im}(\Delta^{\rm cl})$
\end{itemize}
Note that
\begin{align}
    \frac{\mathcal{S}(\mathbb{R}^d)^0}{{\rm im(\Delta^{\rm cl})}}
    \cong
    \frac{\mathcal{S}(\mathbb{R}^d)^0}{{\rm \overline{im(\Delta^{\rm cl})}}},
\end{align}
then it is sufficient to see ${\rm ker}(\pi^0)\cong\overline{{\rm im}(\Delta^{\rm cl})}$.
We show each of them in the later sections.

\subsection{\texorpdfstring{The precise definition of $\pi^0$}{The accurate definition of π0}}
Let $\phi_k$ form an orthogonal basis $B(\mathscr{H})$ and the label $k$ is set in ascending order of the polynomial degree.
Firstly, we define $\tilde{\pi}^0:\mathcal{S}(\mathbb{R}^d)^0 \to \prod_k \operatorname{span} \phi_k \left( = \prod_l \mathscr{H}_l \right)$ as
\begin{equation}
    (\tilde{\pi}^0 (f))_k := \langle \phi_k, f \rangle.
\end{equation}
Then we define $\widetilde{\mathscr{H}}$ as the image of $\tilde{\pi}^0$ and $\pi^0$ as the restriction of $\tilde{\pi}^0$ to $\widetilde{\mathscr{H}}$.

\subsection{\texorpdfstring{The dimension of $\widetilde{\mathscr{H}}$ is at least countably infinite.}{The cardinality of the bases of H is at least countably infinite.}}
In order to show it, we show the existence of $\mathcal{O}_k\in\mathcal{S}(\mathbb{R}^d)$ satisfying
\begin{align}
    (\pi^0(\mathcal{O}_k))_{k^\prime}=
    \begin{cases}
        \phi_k &(k^\prime =k)
        \\
        0 &(k^\prime \neq k).
    \end{cases}
\end{align}
The basic way to check it is the same as the section \ref{subsec:pi^0_surjective}.
Then we set $\mathcal{O}_k$ as
\begin{align}
    \mathcal{O}_k(x_1,\cdots,x_d)
    &=
    \frac{1}{(2\pi)^{d/2}}
    \phi_k(x_1,\cdots,x_d)
    e^{-\frac{1}{2}x^2}\notag\\
    &=
    \frac{1}{(2\pi)^{d/2}}
    \phi_k\left(-\frac{\partial}{\partial x_1},\cdots,-\frac{\partial }{\partial x_d}\right)
    e^{-\frac{1}{2}x^2}.
\end{align}
Thus
\begin{align}
    (\pi^0\left(\mathcal{O}_k\right))_{k^\prime}
    &=
    \left\langle \phi_{k^\prime}, \mathcal{O}_k\right\rangle\notag\\
    &=
    \left\langle \phi_{k^\prime}, 
    \frac{1}{(2\pi)^{d/2}}\phi_k e^{-\frac{1}{2}x^2}\right\rangle\notag\\
    &=
    \delta_{k^\prime,k} \phi_k.
\end{align}
This implies $\pi^0 (\mathcal{O}_k)$ are linear independent and the cardinality of the bases of $\widetilde{\mathscr{H}}$ is at least countably infinite.

Note that $\mathcal{O}_k$ is a natural generalization of $Q$ and $P$ for one-dimensional case in section \ref{subsec:pi^0_surjective}.
\begin{align}
        Q(x)
        &:=
        \frac{1}{\sqrt{\pi}}\exp(x).,\\
        P(x)
        &:=
        -\frac{\partial}{\partial x}Q(x)
        =
        \frac{2}{\sqrt{\pi}}x\exp(x^2).
\end{align}
Hence we can regard $\mathcal{O}_k$ as observables for the asymptotic state in higher dimensions.

\subsection{\texorpdfstring{${\rm ker}(\pi^0)\cong\overline{{\rm im}(\Delta^{\rm cl})}$}{ker π0=closure(im Δ)}}
To show ${\rm ker}(\pi^0)=\overline{{\rm im}(\Delta^{\rm cl})}$, 
we consider the following two steps.
\begin{itemize}
    \item 
    ${\rm ker}(\pi^0)
    \cong 
    \left({\rm ker}(\Delta_{\mathcal{S}'})\right)^\perp$
    \item 
    ${\rm ker}(\Delta_{\mathcal{S}'})
    \cong
    ({\rm im}(\Delta^{\rm cl}))^\perp$
\end{itemize}
 where
\begin{align}
    \left({\rm ker}(\Delta_{\mathcal{S}'})\right)^\perp
    &:=
    \{
    f\in \mathcal{S}(\mathbb{R}^d)\ |\ 
    \forall \phi\in {\rm ker}(\Delta_{\mathcal{S}'}),\ 
    \langle \phi,f\rangle=0
    \},\notag\\
    ({\rm im}(\Delta^{\rm cl}))^\perp
    &:=
    \{
    \phi\in \mathcal{S}'(\mathbb{R}^d)\ |\ 
    \forall f\in {\rm im}(\Delta^{\rm cl}),\ 
    \langle \phi,f\rangle=0
    \}.
\end{align}
By the above equations, we have
\begin{align}
    {\rm ker}(\pi^0)
    \cong
    (({\rm im}(\Delta^{\rm cl}))^\perp)^\perp
    \cong
    \overline{{\rm im}(\Delta^{\rm cl})}.
\end{align}

First of all, we show ${\rm ker}(\pi^0)\cong \left({\rm ker}(\Delta_{\mathcal{S}'})\right)^\perp$.
Take $f\in {\rm ker}(\pi^0)$, 
then for all harmonic polynomial $\phi$ we have
\begin{align}
    \langle \phi, f\rangle=0.
\end{align}
In other words, for all $\phi\in {\rm ker}(\Delta_{\mathcal{S}'})$
\begin{align}
    \langle \phi, f\rangle=0.
\end{align}
Therefore $f\in ({\rm ker}(\Delta_{\mathcal{S}'}))^\perp$, i.e. 
${\rm ker}(\pi^0)\subset \left({\rm ker}(\Delta_{\mathcal{S}'})\right)^\perp$.
By reversing the above discussion, we have ${\rm ker}(\pi^0)\supset \left({\rm ker}(\Delta_{\mathcal{S}'})\right)^\perp$.
Hence, ${\rm ker}(\pi^0)\cong \left({\rm ker}(\Delta_{\mathcal{S}'})\right)^\perp$.

Next, we show ${\rm ker}(\Delta_{\mathcal{S}'})\cong({\rm im}(\Delta^{\rm cl}))^\perp$.
Take $\phi\in {\rm im}(\Delta^{\rm cl}))^\perp$, thus we have
\begin{align}
    &\langle \phi,\Delta f \rangle=0
    \ \ (\forall f\in\mathcal{S}(\mathbb{R}^d))\notag\\
    &\Leftrightarrow
    \langle \Delta_{\mathcal{S}'} \phi,f \rangle=0
    \ \ (\forall f\in\mathcal{S}(\mathbb{R}^d))\notag\\
    &\Leftrightarrow \Delta_{\mathcal{S}'} \phi=0.
\end{align}
Therefore $\phi\in {\rm ker}(\Delta_{\mathcal{S}'})$. Hence ${\rm ker}(\Delta_{\mathcal{S}'})\supset({\rm im}(\Delta^{\rm cl}))^\perp$.
By reversing the above discussion, we have
${\rm ker}(\Delta_{\mathcal{S}'})\subset({\rm im}(\Delta^{\rm cl}))^\perp$. Then we obtain
\begin{align}
    {\rm ker}(\Delta_{\mathcal{S}'})\cong({\rm im}(\Delta^{\rm cl}))^\perp.
\end{align}

\section{Smooth set}
\label{sec:smooth_set}

Most of this section is based on
\cite{Alfonsi:2023qpv}
and 
\cite{Giotopoulos:2023pvl}.

\subsection{Smooth set}
Let $\mathsf{Mfd}$ denote the site of smooth manifolds.
A \textit{smooth set} is a sheaf on $\mathsf{Mfd}$.
In other words, if $\mathsf{Set}$ denotes the category of sets, a smooth set $X$ is a contravariant functor:
\begin{align}
X:\mathsf{Mfd}^{\rm op}\to \mathsf{Set}.
\end{align}
We denote by $\mathsf{SmoothSet} := \mathsf{Sh}(\mathsf{Mfd})$ the category whose objects are smooth sets.
A morphism $\tau : X \to Y$ in this category is given by a natural transformation:
\begin{equation}
\xymatrix{
\mathsf{Mfd}^{\rm op}
\ar@/^18pt/[rr]^-{X}\ar@/_18pt/[rr]_-{Y}\ar@{}[rr]|{\Downarrow^\tau}
&&
\mathsf{Set}
}
\end{equation}

Via the Yoneda embedding $\yo$, a smooth set may be regarded as a certain generalization of a smooth manifold:
\begin{align}
\begin{array}{rcccc}
\yo\colon &\mathsf{Mfd}                     &\longrightarrow& \mathsf{SmoothSet}\\
& \rotatebox{90}{$\in$}&&\rotatebox{90}{$\in$} \\
& M & \longmapsto& \yo(M)(-)
\end{array}
\end{align}
where $\yo(M)(-):=\operatorname{Hom}_{\mathsf{Mfd}}(-,M)$.

A smooth set can handle not only smooth manifolds but also infinite-dimensional spaces such as field configuration spaces.
For example, the configuration space of a real scalar field on a spacetime $M$,
\begin{align}
{\rm Hom}(M,\mathbb{R})
\end{align}
can be regarded as a smooth set by defining
\begin{align}
{\rm ScalarField}(-):={\rm Hom}(M\times -,\mathbb{R})
\end{align}
In this case, a real scalar field $\Phi\in {\rm ScalarField}$ is written as
\begin{align}
\begin{array}{rccc}
\Phi\colon &M\times - &\longrightarrow& \mathbb{R} \\
& \rotatebox{90}{$\in$}&   & \rotatebox{90}{$\in$} \\
& (x,\bullet)  & \longmapsto   & \Phi_\bullet(x)
\end{array}
\end{align}
so that the deformation parameter $\bullet$ accompanies the field $\Phi(x)$.

\subsection{Differential form on smooth set}

A key feature of smooth sets is that one can define differential forms on them.
To prepare for this, let us consider the following smooth set:
\begin{align}
\begin{array}{rccc}
\Omega^p\colon &\mathsf{Mfd}^{\rm op} &\longrightarrow& \mathsf{Set}\\
& \rotatebox{90}{$\in$}&   & \rotatebox{90}{$\in$} \\
& -  & \longmapsto & \Omega^p(-)
\end{array}
\end{align}
Here, the exterior derivative
${\rm d}:\Omega^p(-)\to \Omega^{p+1}(-)$
induces a morphism of $\mathsf{SmoothSet}$
\begin{align}
{\rm d}:\Omega^p\to \Omega^{p+1}.
\end{align}
Now, the space of $p$-forms on a smooth set
$X$ is defined by
\begin{align}
\bm{\Omega}^p(X):={\rm Hom}_{\mathsf{SmoothSet}}(X,\Omega^p).
\end{align}
Note that $\bm{\Omega}^p(X)$ is not an object of $\mathsf{SmoothSet}$, but rather a hom-set.
By the Yoneda lemma, for any smooth manifold $M$,
\begin{align}
\bm{\Omega}^p(\yo(M))\cong \Omega^p(M)
\end{align}
so this definition is a natural generalization of the usual notion of differential forms.

Moreover, if $\omega^{(p)}\in \bm{\Omega}^{p}(X)$, then it is a morphism
\begin{align}
\omega^{(p)}:X\to \Omega^p,
\end{align}
so we can compose it with the exterior derivative
${\rm d}:\Omega^p\to\Omega^{p+1}$ 
to define
\begin{align}
{\rm d} \omega^{(p)}
:=
{\rm d}\circ \omega^{(p)}
\in 
\bm{\Omega}^{p+1}(X)
\end{align}
This satisfies ${\rm d}^2=0$.
Also, note that by definition we have
$\Omega^0=\yo(\mathbb{R}).$

\subsection{Example of application: variation of the action functional}

In this subsection, we reexamine the computation of the variation of the action by using differential forms on smooth sets.
This makes the physical meaning of the deformation parameter $\bullet$ of the field $\Phi_\bullet(x)$ transparent.

Let us consider an action functional $S$
\begin{align}
\begin{array}{rccc}
S\colon 
&{\rm Hom}(M,\mathbb{R})
&\longrightarrow
&\mathbb{R}
\\
&\rotatebox{90}{$\in$}
&   
&\rotatebox{90}{$\in$} 
\\
&\Phi& \longmapsto   
&S(\Phi).
\end{array}
\end{align}
By replacing 
$\Phi\in {\rm Hom}(M,\mathbb{R})$ 
with 
$\Phi\in {\rm ScalarField}$, 
we obtain
\begin{align}
\begin{array}{rccc}
S\colon 
&{\rm ScalarField}
&\longrightarrow
&\yo(\mathbb{R})=\Omega^0
\\
&\rotatebox{90}{$\in$}
&   
&\rotatebox{90}{$\in$} 
\\
&\Phi_\bullet
& \longmapsto   
&S(\Phi_\bullet).
\end{array}
\end{align}
Hence, we have
$S\in \bm{\Omega}^0({\rm ScalarField})$.

Applying the exterior derivative to this, we obtain
\begin{align}
\begin{array}{rccc}
{\rm d}\colon 
&\Omega^0
&\longrightarrow
&\Omega^1
\\
&\rotatebox{90}{$\in$}
&   
&\rotatebox{90}{$\in$} 
\\
&S(\Phi_\bullet)
& \longmapsto   
&{\rm d}S(\Phi_\bullet).
\end{array}
\end{align}

In fact, the above construction corresponds to the usual variation of the action.
As a concrete example, let us consider a massless scalar action
\begin{align}
S(\Phi_\bullet)
=
\frac{1}{2}
\int_{x\in M}
\left(
\frac{\partial}{\partial x}\Phi_\bullet(x)
\right)^2
\in\Omega^0(-)
\end{align}
where $\bullet\in -$. 
Applying the exterior derivative yields
\begin{align}
{\rm d}S(\Phi_\bullet)
=
\int_{x\in M}
\left(
\frac{\partial}{\partial x}
\Phi_\bullet(x)
\right)
\frac{\partial}{\partial x}
{\rm d}\Phi_\bullet(x)
,\ 
{\rm d}\Phi_\bullet(x)
=
\sum_{i=1}^{m}
\left(
\frac{\partial}{\partial\bullet^i}
\Phi_\bullet(x)
\right)
{\rm d}\bullet^i.
\end{align}
Here $m$ denotes the dimension of $-$.
If $M$ is closed, 
\begin{align}
{\rm d}S(\Phi_\bullet)
=
\int_{x\in M}
\left(
-
\left(
\frac{\partial}{\partial x}
\right)^2
\Phi_\bullet(x)
\right)
{\rm d}\Phi_\bullet(x),
\end{align}
then this is the same as the usual variation of the action.

From the above computation, it is clear that the parameter $\bullet\in -$ corresponds to the degrees of freedom that deform the scalar field $\Phi$ in various ways.
The smoothness of a function on a smooth set is defined with respect to this deformation parameter. 
Similarly, in the case of a state
\begin{align}
\langle-\rangle:H^0{\rm Obs}(M)\to\mathbb{C},
\end{align}
we regard $H^0{\rm Obs}(M)$ as a smooth set, and in this way we can define the smoothness of the state.

\bibliographystyle{ptephy}
\bibliography{sample}

\end{document}